# New approach for Damping in a squeezed bath and its time evolution through Complete Class of Gaussian Quasi-distributions

Mohammad Reza Bazrafkan, Seyed Mahmoud Ashrafi[*] and Fahimeh Naghdi

Physics Department, Faculty of Science, I. K. I. University, Qazvin, I. R. Iran

By virtue of the thermo-entangled states representation of density operator and using "dissipative interaction picture" we solve the master equation of a driven damped harmonic oscillator in a squeezed bath. We show that the essential part of the dynamics can be expressed by the convolution of initial Wigner function with a special kind of normalized Gaussian in phase space and relate the dynamics with the change of Gaussian ordering of density operator.

## 1-Introduction

Dissipative interactions in the presence of a heat bath in squeezed vacuum state can be analyzed using phase space representations of density operator [1-4]. In this work, instead of using these conventional techniques we treat the time evolution of density operator by virtue of the entangled state representation [5, 6] and also employ "dissipative interaction picture" to solve the problem. The method of entangled state representation for solving master equation of an open system is a newly developed method, and introduced and widely used by H.Y. Fan [7-9]. In this representation one can represent any single mode operator, for example density operator, by a vector in a two-mode Fock space of which second mode is a fictitious mode. In this way master equation appears as a Schrödinger-like time evolution equation and therefore familiar techniques, for example interaction picture, can be used in a new content.

The aim of this work is to solve the master equation of a damped harmonic oscillator which is externally driven by a classical force field and immersed in a squeezed bath. As a new method we introduce "dissipative interaction picture" for entangled state representation of density operator. Moreover we relate time evolution processes to different quasi-distribution representation of initial Wigner function with time dependent ordering parameter.

This paper is organized as follows.
In Sec. 2, we briefly review some important elements of thermo-entangled state representation. In Sec. 3, we review those concepts from complete Gaussian class of quasi-distributions which are needed for our discussions. In Sec. 4, we solve the problem of driven damped harmonic oscillator in squeezed bath using "dissipative interaction picture" in entangled state representation. Sec. 5 is devoted to calculation of Wigner function. In Sec. 6 we relate time evolution of the Wigner function to complete calss of quasi-distributions. Sec. 7 is devoted to conclusions.

## 2- A brief review of the thermo entangled state representation

Consider a two-mode Fock space and let $(\hat{a}, \hat{a}^\dagger)$ and $(\hat{b}, \hat{b}^\dagger)$ be the pairs of canonical annihilation and creation operators of the first (physical) and the second (fictitious) mode respectively. Defining $\hat{\eta} \equiv \hat{a} - \hat{b}^\dagger$ and $\hat{\eta}^\dagger = \hat{a}^\dagger - \hat{b}$ one can recognize that $[\hat{\eta}, \hat{\eta}^\dagger] = 0$ and therefore there exist a set of common eigen-vectors of $(\hat{\eta}, \hat{\eta}^\dagger)$ called thermo-entangled states, and defined by

$$\hat{\eta}|\eta\rangle_\eta = \eta|\eta\rangle_\eta, \quad \hat{\eta}^\dagger|\eta\rangle_\eta = \eta^*|\eta\rangle_\eta, \quad \eta \in \mathbb{C}. \tag{1}$$

---

[*] Corresponding author  ashrafi@alm.ikiu.ac.ir

In [10, 11] it is shown that these states can be given by

$$|\eta\rangle_\eta = \exp\left(-\frac{|\eta|^2}{2} + \eta a^\dagger - \eta^* b^\dagger + a^\dagger b^\dagger\right)|0,0\rangle_{\text{coh.}}, \qquad \eta \in \mathbb{C}, \tag{2}$$

Where $|0,0\rangle_{\text{coh.}}$ is two-mode vacuum state and Thermo entangled states obey orthogonality condition and provide the following decomposition of identity operator of the doubled Fock space

$$_\eta\langle\eta'|\eta''\rangle_\eta = \pi\delta^{(2)}(\eta' - \eta''), \qquad \int \frac{d^2\eta}{\pi}|\eta\rangle_\eta{}_\eta\langle\eta| = \hat{1}. \tag{3}$$

Thermo entangled states can be used to "represent" any operator $\hat{\rho}$ of the physical mode by a ket-vector $|\rho\rangle \equiv \hat{\rho}|0\rangle_\eta$ of the doubled Fock space. Moreover, if $|\rho\rangle$ is given then there exist a unique $\hat{\rho}$ (related to physical mode) which has this thermo-entangled state representation. If $\hat{\rho}$ is a density operator then ket-vector $|\rho\rangle$ contains all information about quantum state. For example expectation value of an operator $\hat{A}$ may be found as follows

$$\left\langle \hat{A} \right\rangle_\rho = \text{Tr}\left\{\hat{A}\hat{\rho}\right\} = {}_\eta\langle\eta = 0|\hat{A}\hat{\rho}|\eta = 0\rangle_\eta = \left\langle A^\dagger \big| \rho \right\rangle. \tag{4}$$

Using thermo-entangled state representation of density operator one can recast master equation of an open system to a new Schrödinger-like equation and solve it. A useful property of thermo-entangled state representation of density operator is that we can directly calculate related Wigner function from it. It is shown that [5, 6]

$$|\eta\rangle_\eta = \hat{D}_a(\eta)|0\rangle_\eta \equiv e^{\eta\hat{a}^\dagger - \eta^*\hat{a}}|0\rangle_\eta, \tag{5}$$

Using Eq. (4) and Eq. (5) the Wigner function $W(\alpha)$ can be find as follows

$$\begin{aligned} W(\alpha) &= \int \frac{d^2\eta}{\pi} e^{\alpha\eta^* - \alpha^*\eta} \text{Tr}\{\hat{D}_a(\eta)\hat{\rho}\}, \\ &= \int \frac{d^2\eta}{\pi} e^{\alpha\eta^* - \alpha^*\eta} {}_\eta\langle 0|\hat{D}_a(\eta)\hat{\rho}|0\rangle_\eta, \\ &= \int \frac{d^2\eta}{\pi} e^{\alpha\eta^* - \alpha^*\eta} {}_\eta\langle -\eta|\rho\rangle. \end{aligned} \tag{6}$$

This means that ${}_\eta\langle -\eta|\rho\rangle$ and $W(\alpha)$ relates to each other by Fourier transform.

## 3- The complete Gaussian class of quasi-distributions

Let us briefly review some basic concepts of Guassian quasi-distribution representation of quantum states [12]. Every $\vec{r}$-parameterized quasi-distribution for density operator $\hat{\rho}$ corresponds to transition operator $\hat{T}_{\vec{r}}(\alpha)$, $\alpha \in \mathbb{C}$, which is defined by a convolution relation as follows

$$\hat{T}_{\vec{r}}(\alpha) = g_{\vec{r}}(\alpha) * \hat{T}_{\vec{0}}(\alpha) \triangleq \int \frac{d^2\beta}{\pi} g_{\vec{r}}(\beta)\hat{T}_{\vec{0}}(\alpha - \beta). \tag{7}$$

Here $\hat{T}_{\vec{0}}(\alpha)$ is the Wigner operator, i.e. Fourier transform of displacement operator

$$\hat{T}_{\vec{0}}(\alpha) = \int \frac{d^2\beta}{\pi} e^{\alpha\beta^* - \alpha^*\beta} e^{\beta\hat{a}^\dagger - \beta^*\hat{a}}, \qquad \alpha \in \mathbb{C}, \tag{8}$$

and $g_{\vec{r}}(\alpha)$, $\vec{r} \in \mathbb{C}^3$, is a normalized function on complex plane (phase space) defined by

$$g_{\vec{r}}(\alpha) = \frac{2}{\sqrt{(\vec{r}\cdot\vec{r})}} \exp\left(-\frac{1}{(\vec{r}\cdot\vec{r})}\left\{r_1(\alpha^2 - \alpha^{*2}) + ir_2(\alpha^2 + \alpha^{*2}) + r_3 2\alpha\alpha^*\right\}\right). \tag{9}$$

If $\hat{\rho}$ is density operator of a quantum state then related $\vec{r}$-parameterized quasi distribution is defined by

$$W_{\vec{r}}(\alpha) = \text{Tr}\{\hat{T}_{\vec{r}}(\alpha)\hat{\rho}\} = g_{\vec{r}}(\alpha) * \text{Tr}\{\hat{T}_{\vec{0}}(\alpha)\hat{\rho}\} = g_{\vec{r}}(\alpha) * W_{\vec{0}}(\alpha). \tag{10}$$

The set of transition operators obey following completeness and orthogonality relation [12]

$$\int \frac{d^2\alpha}{\pi}\hat{T}_{\vec{r}}(\alpha) = \hat{1}, \qquad \left\langle \hat{T}_{\vec{r}}(\alpha)\hat{T}_{-\vec{r}}(\beta) \right\rangle = \pi\delta^{(2)}(\alpha - \beta), \tag{11}$$

and density operator can be reconstructed as
$$\hat{\rho} = \int \frac{d^2\alpha}{\pi} W_{\vec{r}}(\alpha) \hat{T}_{-\vec{r}}(\alpha). \quad (12)$$

The concept of Guassian quasi-distributions closely related to Gaussian class of orderings with vector parameter $\vec{r} \in \mathbb{C}^3$. In [13] ordering "operator" $\{\cdots\}_{\vec{r}}$ is defined by
$$\{e^{\xi\hat{a}^\dagger - \xi^*\hat{a}}\}_{\vec{r}} \triangleq \tilde{g}_{\vec{r}}(\xi, \xi^*) e^{\xi\hat{a}^\dagger - \xi^*\hat{a}}, \qquad \{\hat{a}^{\dagger m}\hat{a}^n\}_{\vec{r}} = \partial_\xi^m \partial_{-\xi^*}^n \left[v(\xi,\xi^*)e^{\xi\hat{a}^\dagger - \xi^*\hat{a}}\right]_{\xi=\xi^*=0}, \quad (13)$$

Where $\tilde{g}_{\vec{r}}(\xi, \xi^*)$ is Fourier transform of Eq. (9). Fundamental property of $\hat{R} = \{\hat{a}^{\dagger m}\hat{a}^n\}_{-\vec{r}}$ is that its $\vec{r}$ parameterized quasi-distribution representation can be found by following simple rule
$$W_{\vec{r}}^R(\alpha) = \mathrm{Tr}\{\hat{T}_{\vec{r}}(\alpha)\{\hat{a}^{\dagger m}\hat{a}^n\}_{-\vec{r}}\} = \alpha^{*m}\alpha^n. \quad (14)$$

## 4 - Driven harmonic oscillator in a squeezed bath

Consider a quantum harmonic oscillator with natural frequency $\Omega$ which is driving by time dependent uniform force field and let
$$\hat{H} = \hbar\Omega\hat{a}^\dagger\hat{a} + \hbar f(t)(\hat{a} + \hat{a}^\dagger), \quad (15)$$
be the Hamiltonian operator. If this system has interaction with a heat bath in squeezed vacuum state then its master equation for density operator in Schrödinger picture is [1,2]
$$\frac{d\hat{\rho}_s(t)}{dt} = \frac{1}{i\hbar}[\hat{H}, \hat{\rho}_s] + \kappa(\bar{n}+1)(2\hat{a}\hat{\rho}_s\hat{a}^\dagger - \hat{a}^\dagger\hat{a}\hat{\rho}_s - \hat{\rho}_s\hat{a}^\dagger\hat{a}) + \kappa\bar{n}(2\hat{a}^\dagger\hat{\rho}_s\hat{a} - \hat{a}\hat{a}^\dagger\hat{\rho}_s - \hat{\rho}_s\hat{a}\hat{a}^\dagger)\cdots$$
$$\cdots + \kappa M(2\hat{a}^\dagger\hat{\rho}_s\hat{a}^\dagger - \hat{a}^\dagger\hat{a}^\dagger\hat{\rho}_s - \hat{\rho}_s\hat{a}^\dagger\hat{a}^\dagger) + \kappa M^*(2\hat{a}\hat{\rho}_s\hat{a} - \hat{a}\hat{a}\hat{\rho}_s - \hat{\rho}_s\hat{a}\hat{a}). \quad (16)$$

Where the coefficients $\bar{n}$ and $\kappa$ are the average photon number of the thermal bath and the damping constant respectively, and $M$ is the parameter of a squeezed reservoir. Using change of variable $\hat{\rho}(t) = e^{i\Omega t\hat{a}^\dagger\hat{a}}\hat{\rho}_s(t)e^{-i\Omega t\hat{a}^\dagger\hat{a}}$, which may be called "rotational interaction picture", we find a more simplified form of the master equation as follows
$$\frac{d\hat{\rho}(t)}{dt} = -if(t)[(\hat{a}e^{-i\Omega t} + \hat{a}^\dagger e^{i\Omega t}), \hat{\rho}] +$$
$$\cdots \kappa(\bar{n}+1)(2\hat{a}\hat{\rho}\hat{a}^\dagger - \hat{a}^\dagger\hat{a}\hat{\rho} - \hat{\rho}\hat{a}^\dagger\hat{a}) + \bar{n}\kappa(2\hat{a}^\dagger\hat{\rho}\hat{a} - \hat{a}\hat{a}^\dagger\hat{\rho} - \hat{\rho}\hat{a}\hat{a}^\dagger)\cdots \quad (17)$$
$$\ldots + \kappa Me^{2i\Omega t}(2\hat{a}^\dagger\hat{\rho}\hat{a}^\dagger - \hat{a}^\dagger\hat{a}^\dagger\hat{\rho} - \hat{\rho}\hat{a}^\dagger\hat{a}^\dagger) + \kappa M^*e^{-2i\Omega t}(2\hat{a}\hat{\rho}\hat{a} - \hat{a}\hat{a}\hat{\rho} - \hat{\rho}\hat{a}\hat{a}).$$

In view of definition Eq. (1) for $\eta = 0$ and by acting both sides of Eq. (17) on $|0\rangle_\eta$ one can find a Schrödinger like time evolution
$$\frac{d|\rho(t)\rangle}{dt} = \{\hat{G}_0 + \hat{g}(t)\}|\rho(t)\rangle, \quad (18)$$
Where
$$\hat{G}_0 = \kappa(\bar{n}+1)(\hat{\eta}\hat{b} - \hat{\eta}^\dagger\hat{a}) + \kappa\bar{n}(\hat{\eta}^\dagger\hat{b}^\dagger - \hat{\eta}\hat{a}^\dagger),$$
$$\hat{g}(t) = -if(t)(\hat{\eta}e^{-i\Omega t} + \hat{\eta}^\dagger e^{i\Omega t}) - \kappa Me^{2i\Omega t}\hat{\eta}^{\dagger 2} - \kappa M^*e^{-2i\Omega t}\hat{\eta}^2. \quad (19)$$

It is a simple task to check following commutation relations
$$[\hat{G}_0, \hat{\eta}] = -\kappa\hat{\eta}, \qquad [\hat{G}_0, \hat{\eta}^\dagger] = -\kappa\hat{\eta}^\dagger. \quad (20)$$

Using these relations one can find that $[\hat{G}_0 + \hat{g}(t), \hat{G}_0 + \hat{g}(t')] \neq 0$, therefore integration of Eq. (18) is not straightforward. To solve Eq. (18) we introduce following "dissipative interaction picture" defined by
$$|\rho(t)\rangle_I = e^{-\hat{G}_0 t}|\rho(t)\rangle, \quad (21)$$
for entangled state representation of density operator. For new variable $|\rho(t)\rangle_I$ we have
$$\frac{d|\rho(t)\rangle_I}{dt} = \hat{g}_I(t)|\rho(t)\rangle_I, \qquad \hat{g}_I(t) = e^{-\hat{G}_0 t}\hat{g}(t)e^{\hat{G}_0 t}. \quad (22)$$

Making use of commutation relations Eq. (20) one can find

$$\hat{g}_I(t) = -\left\{ if(t)\left(\hat{\eta}e^{-i\Omega t}e^{\kappa t} + \hat{\eta}^\dagger e^{i\Omega t}e^{\kappa t}\right) + \kappa M e^{2i\Omega t}\hat{\eta}^{\dagger 2}e^{2\kappa t} + \kappa M^* e^{-2i\Omega t}e^{2\kappa t}\hat{\eta}^2 \right\}. \tag{23}$$

Now because of $\left[\hat{g}_I(t), \hat{g}_I(t')\right] = 0$ we can easily integrate Eq. (22) and find

$$|\rho(t)\rangle_I = \exp\left\{\int_0^t dt' \hat{g}_I(t')\right\}|\rho_s(0)\rangle, \tag{24}$$

where we use initial condition $|\rho(0)\rangle_I = |\rho(0)\rangle = |\rho_s(0)\rangle$. Solving the above integral leads to the following solution

$$|\rho(t)\rangle_I = e^{\lambda_1(t)\hat{\eta}^\dagger - \lambda_1^*(t)\hat{\eta}} e^{\lambda_2(t)\hat{\eta}^{\dagger 2} + \lambda_2^*(t)\hat{\eta}^2}|\rho_s(0)\rangle, \tag{25}$$

where

$$\lambda_1(t) \triangleq -i\int_0^t dt' f(t') e^{\kappa t' + i\Omega t'}, \qquad \lambda_2(t) \triangleq -\kappa M \int_0^t dt' e^{2\kappa t' + i2\Omega t'}. \tag{26}$$

In view of identities $\hat{a}|0\rangle_\eta = \hat{b}^\dagger|0\rangle_\eta$ and $\hat{a}^\dagger|0\rangle_\eta = \hat{b}|0\rangle_\eta$ it is straightforward to show that

$$e^{\lambda_2\hat{\eta}^{\dagger 2} + \lambda_2^*\hat{\eta}^2}\rho_s(0)|0\rangle_\eta = \sum_{m,n} \frac{\lambda_2^m \lambda_2^{*n}(-2)^{m+n}}{m!n!} \hat{a}^{\dagger m} e^{\lambda_2 \hat{a}^{\dagger 2}} \hat{a}^n e^{\lambda_2^* \hat{a}^2} \rho_s(0) e^{\lambda_2^* \hat{a}^2} \hat{a}^n e^{\lambda_2 \hat{a}^{\dagger 2}} \hat{a}^{\dagger m}|0\rangle_\eta. \tag{27}$$

Now because of identity $e^{\lambda\hat{\eta}^\dagger - \lambda^*\hat{\eta}} = \hat{D}_a(\lambda)\hat{D}_b(\lambda^*)$, we can rewrite the solution Eq. (24) as follows

$$\begin{aligned}|\rho(t)\rangle_I &= \sum_{m,n=0}^\infty \frac{\lambda_2^m \lambda_2^{*n}(-2)^{m+n}}{m!n!} \hat{D}_a(\lambda_1) \hat{a}^{\dagger m} e^{\lambda_2 \hat{a}^{\dagger 2}} \hat{a}^n e^{\lambda_2^* \hat{a}^2} \rho_s(0) e^{\lambda_2^* \hat{a}^2} \hat{a}^n e^{\lambda_2 \hat{a}^{\dagger 2}} \hat{a}^{\dagger m} \hat{D}_a^\dagger(\lambda_1)|0\rangle_\eta, \\ &\equiv \hat{R}|0\rangle_\eta,\end{aligned} \tag{28}$$

In which $\hat{R}$ is an operator related to physical mode. To find the density operator in rotational interaction picture, i.e. $|\rho(t)\rangle = e^{\hat{G}_0 t}|\rho(t)\rangle_I$, after some algebra we can write

$$\begin{aligned}|\rho(t)\rangle &= e^{\hat{G}_0 t}\hat{R}|0\rangle_\eta, \\ &= \exp\left\{-2\kappa t\bar{n} - \kappa t(2\bar{n}+1)\left(\hat{a}^\dagger\hat{a} + \hat{b}^\dagger\hat{b}\right) + 2\kappa t\bar{n}\hat{a}^\dagger\hat{b}^\dagger + 2\kappa t(\bar{n}+1)\hat{a}\hat{b}\right\}\hat{R}|0\rangle_\eta.\end{aligned} \tag{29}$$

Now, we must rewrite right hand side of the above equation as the action of a unique density operator $\hat{\rho}$ related to physical mode on $|0\rangle_\eta$. To this end we employ following operator identity [14]

$$\begin{aligned}&\exp\left\{h\left(\hat{a}^\dagger\hat{a} + \hat{b}^\dagger\hat{b}\right) + g\hat{a}^\dagger\hat{b}^\dagger + r\hat{a}\hat{b}\right\} = \\ &e^{-h}\exp\left\{\frac{g}{D\coth D - h}\hat{a}^\dagger\hat{b}^\dagger\right\}\exp\left\{\ln\frac{D\operatorname{sech}D}{D - h\tanh D}\left(\hat{a}^\dagger\hat{a} + \hat{b}^\dagger\hat{b} + 1\right)\right\}\exp\left\{\frac{r}{D\coth D - h}\hat{a}\hat{b}\right\},\end{aligned} \tag{30}$$

in which $D^2 = h^2 - gr$ and find

$$|\rho(t)\rangle = e^{\kappa t}\exp\left[2\bar{n}A\sinh(\kappa t)\hat{a}^\dagger\hat{b}^\dagger\right]\exp\left[\ln(A)\left(\hat{a}^\dagger\hat{a} + \hat{b}^\dagger\hat{b} + 1\right)\right]\exp\left[2(\bar{n}+1)A\sinh(\kappa t)\hat{a}\hat{b}\right]\hat{R}|0\rangle_\eta, \tag{31}$$

Where, time dependent coefficient $A(t)$ defined as follows

$$A\sinh\kappa t \triangleq \frac{1}{\coth\kappa t + 2\bar{n} + 1}. \tag{32}$$

After some simple algebra one can find that

$$2A\sinh\kappa t = \frac{T}{\bar{n}T + 1}, \quad Ae^{\kappa t} = \frac{1}{\bar{n}T + 1}, \tag{33}$$

Where $T \equiv 1 - e^{-2\kappa t}$. Now, expanding exponential functions in Eq. (31) leads to

$$\hat{\rho}(t) = \frac{1}{\bar{n}T + 1}\sum_{p,q=0}^\infty \frac{1}{p!q!}\left(\frac{\bar{n}T}{\bar{n}T + 1}\right)^p\left(\frac{(\bar{n}+1)T}{\bar{n}T + 1}\right)^q \hat{a}^{\dagger p} e^{\ln(A)\hat{a}^\dagger\hat{a}} \hat{a}^q \hat{R}(t) \hat{a}^{\dagger q} e^{\ln(A)\hat{a}^\dagger\hat{a}} \hat{a}^p, \tag{34}$$

where we use identities $\hat{a}|0\rangle_\eta = \hat{b}^\dagger|0\rangle_\eta$ and $\hat{a}^\dagger|0\rangle_\eta = \hat{b}|0\rangle_\eta$. This formula is the general solution of the problem of driven damped quantum harmonic oscillator in squeezed bath.

# 5 - Wigner function calculation

In this section we find Wigner function representation of formal solution Eq. (34). In view of Eq. (6) one can find Wigner function representation as follows. We begin by calculating ${}_\eta\langle\eta|e^{\hat{G}_0 t}$ ,

$$_\eta\langle\eta|e^{\hat{G}_0 t} = {}_\eta\langle\eta|e^{\kappa t\{(\hat{a}\hat{b}-\hat{a}^\dagger\hat{b}^\dagger+1)-(2\bar{n}+1)\hat{\eta}^\dagger\hat{\eta}\}}. \tag{35}$$

Due to commutation relation $\left[(\hat{a}\hat{b}-\hat{a}^\dagger\hat{b}^\dagger+1),\hat{\eta}\hat{\eta}^\dagger\right]=-2\hat{\eta}\hat{\eta}^\dagger$ and using operator identity [15]

$$\left[\hat{A},\hat{B}\right]=\tau\hat{B} \quad\mapsto\quad e^{\lambda(\hat{A}+\sigma\hat{B})}=e^{\lambda\hat{A}}e^{\frac{\sigma}{\tau}(1-e^{-\lambda\tau})\hat{B}}, \tag{36}$$

One can find

$$_\eta\langle\eta|e^{\hat{G}_0 t} = e^{\kappa t}\,{}_\eta\langle\eta|e^{\kappa t(\hat{a}\hat{b}-\hat{a}^\dagger\hat{b}^\dagger)}e^{\frac{(2\bar{n}+1)}{2}(1-e^{2\kappa t})\hat{\eta}\hat{\eta}^\dagger}. \tag{37}$$

Now because of ${}_\eta\langle\eta|e^{\lambda(\hat{a}\hat{b}-\hat{a}^\dagger\hat{b}^\dagger)}=e^{-\lambda}\,{}_\eta\langle\eta e^{-\lambda}|$, which is proved in [5, 6], we find desired expression as follows

$$_\eta\langle\eta|e^{\hat{G}_0 t} = e^{-(\bar{n}+1/2)T|\eta|^2}\,{}_\eta\langle\eta e^{-\kappa t}|, \qquad T(t)\equiv 1-e^{-2\kappa t}. \tag{38}$$

To employ Eq. (6) for calculation Wigner function we need ${}_\eta\langle\eta|\rho(t)\rangle$, and from Eq. (18, 25) we have

$$\begin{aligned}_\eta\langle\eta|\rho(t)\rangle &= {}_\eta\langle\eta|e^{\hat{G}_0 t}|\rho(t)\rangle_I, \\ &= e^{-(\bar{n}+1/2)T|\eta|^2}\,{}_\eta\langle\eta e^{-\kappa t}|e^{\lambda_1\hat{\eta}^\dagger-\lambda_1^*\hat{\eta}+\lambda_2\hat{\eta}^{\dagger 2}+\lambda_2^*\hat{\eta}^2}|\rho_s(0)\rangle, \\ &= e^{-(\bar{n}+1/2)T|\eta|^2+\lambda_1 e^{-\kappa t}\eta^*-\lambda_1^* e^{-\kappa t}\eta+\lambda_2 e^{-2\kappa t}\eta^{*2}+\lambda_2^* e^{-2\kappa t}\eta^2}\,{}_\eta\langle\eta e^{-\kappa t}|\rho_s(0)\rangle. \end{aligned} \tag{39}$$

Finally equation Eq. (6) leads to

$$W(\alpha,t)=\int\frac{d^2\eta}{\pi}e^{\alpha^*\eta-\alpha\eta^*}e^{-(\bar{n}+1/2)T|\eta|^2+\lambda_2 e^{-2\kappa t}\eta^{*2}+\lambda_2^* e^{-2\kappa t}\eta^2+\lambda_1 e^{-\kappa t}\eta^*-\lambda_1^* e^{-\kappa t}\eta}\,{}_\eta\langle\eta e^{-\kappa t}|\rho_s(0)\rangle. \tag{40}$$

To simplify this result, we rewrite above relation as follows

$$e^{-2\kappa t}W\left(e^{-\kappa t}\alpha,t\right)=\int\frac{d^2\beta}{\pi}\frac{d^2\eta}{\pi}e^{-(\bar{n}+1/2)e^{2\kappa t}T|\eta|^2+\lambda_2^*\eta^2+\lambda_2\eta^{*2}+(\alpha-\beta-\lambda_1)^*\eta-(\alpha-\beta-\lambda_1)\eta^*}W(\beta,0), \tag{41}$$

Where in the last step, inverse Fourier transform of Eq. (6) is used. Further simplification can be made if one uses change of variable $\beta'=\alpha-\beta-\lambda_1$ and integration formula

$$\int\frac{d^2 z}{\pi}\exp\left\{\varsigma|z|^2+\xi z+\eta z^*+fz^2+gz^{*2}\right\}=\frac{1}{\sqrt{\varsigma^2-4fg}}\exp\left[\frac{-\varsigma\xi\eta+\xi^2 g+\eta^2 f}{\varsigma^2-4fg}\right], \tag{42}$$

Therefore leads to

$$e^{-2\kappa t}W\left(e^{-\kappa t}(\alpha+\lambda_1),t\right)=\int\frac{d^2\beta}{\pi}g(\beta,t)W(\alpha-\beta,0)\triangleq g(\alpha,t)*W(\alpha,0). \tag{43}$$

Right hand side of this relation explicitly is the convolution of normalized function $g(\beta,t)$ with initial Wigner function, where

$$g(\beta,t)=\frac{2}{\sqrt{(2\bar{n}+1)^2 e^{4\kappa t}T^2-|4\lambda_2|^2}}\exp\left\{-\frac{2(2\bar{n}+1)e^{2\kappa t}T\beta\beta^*-4\lambda_2^*\beta^2-4\lambda_2\beta^{*2}}{(2\bar{n}+1)^2 e^{4\kappa t}T^2-|4\lambda_2|^2}\right\}. \tag{44}$$

# 6 – Relation to complete class of Gaussian quasi distributions

In this section we show that, how the essential part of time evolution Eq. (43) relates to change of ordering parameter $\vec{r}$ in Gaussian quasi-distribution representation of density operator. Direct integration show that $g(\beta,t)$ in Eq. (44) is normalized i.e.

$$\int\frac{d^2\beta}{\pi}g(\beta,t)=1, \tag{45}$$

and comparing with Eq. (9) show that $g(\alpha,t)=g_{\vec{r}(t)}(\alpha)$ where time dependent ordering parameter $\vec{r}$ is

$$r_1(t) = i4\operatorname{Im}(\lambda_2), \quad r_2(t) = i\operatorname{Re}(\lambda_2), \quad r_3(t) = (2\bar{n}+1)(e^{2\kappa t}-1). \tag{46}$$

If we define $W'(\alpha,t) = e^{-2\kappa t}W(e^{-\kappa t}(\alpha+\lambda_1),t)$ as a map which explicitly conserve normalization, we have

$$W'(\alpha,t) = e^{-2\kappa t}W(e^{-\kappa t}(\alpha+\lambda_1),t) = g_{\vec{r}(t)}(\alpha) * W(\alpha,0). \tag{47}$$

For example if initial state is $W(\alpha,0) = g_{\vec{s}}(\alpha)$, Because of the property $g_{\vec{r}}(\alpha) * g_{\vec{s}}(\alpha) = g_{\vec{r}+\vec{s}}(\alpha)$ we can write [12]

$$e^{-2\kappa t}W(e^{-\kappa t}(\alpha+\lambda_1),t) = g_{\vec{r}(t)}(\alpha) * g_{\vec{s}}(\alpha) = g_{\vec{s}+\vec{r}(t)}(\alpha), \quad \Rightarrow$$
$$W(\alpha,t) = e^{2\kappa t}g_{\vec{s}+\vec{r}(t)}(e^{\kappa t}\alpha - \lambda_1). \tag{48}$$

As another example consider coherent state $\hat{\rho}(0) = |\alpha_0\rangle\langle\alpha_0|$ as initial state with Wigner function

$$W(\alpha,0) = 2\exp\{-2|\alpha-\alpha_0|^2\} = \pi\delta(\alpha-\alpha_0) * 2\exp\{-2|\alpha|^2\},$$
$$= \pi\delta(\alpha-\alpha_0) * g_{(0,0,1)}(\alpha). \tag{49}$$

In view of (14) time evolution of the Wigner function is

$$e^{-2\kappa t}W(e^{-\kappa t}(\alpha+\lambda_1),t) = \pi\delta(\alpha-\alpha_0) * g_{\vec{r}(t)}(\alpha) * g_{(0,0,1)}(\alpha),$$
$$= \pi\delta(\alpha-\alpha_0) * g_{(r_1,r_2,1+r_3)}(\alpha) = g_{(r_1,r_2,1+r_3)}(\alpha-\alpha_0), \tag{50}$$

In the last step we have used the fact that $\pi\delta(\alpha-\alpha_0) * \cdots$ act as a translation operator. Finally we have

$$W(\alpha,t) = e^{2\kappa t}g_{(r_1,r_2,1+r_3)}(e^{\kappa t}\alpha - \alpha_0 - \lambda_1). \tag{51}$$

In more general case every normalized function on complex plane (phase space) can be written as successive convolutions of the following functions [12]

$$W'(\alpha) = \pi\delta^{(2)}(\alpha-\alpha_0) * g_{\vec{s}}(\alpha) * \cdots \tag{52}$$

In which $\alpha_0$ is a complex number and $\vec{s} \in \mathbb{C}^3$ is a complex vector and so on. Convolution with $g_{\vec{r}(t)}(\alpha)$ affect only on the second factor of the above resolution and other parameters remain unchanged. In fact the set of normalized functions armed with convolution product forms a commutative group and convolution with $g_{\vec{r}}(\alpha)$, $\vec{r} \in \mathbb{C}^3$ is a subgroup which physically realized when our system interacts with a squeezed bath.

## 7- Conclusions

In summary, we have employed "dissipative interaction picture" in entangled state representation of density operator for treating the time evolution of the quantum state of a driven harmonic oscillator in squeezed heat bath. Moreover we show that the essential part of the dynamics can be expressed by the convolution of initial Wigner function with function $g_{\vec{r}(t)}(\alpha)$ related to complete class of Gaussian quasi-distributions.